# Multi-photon absorption properties of semiconducting nanomaterials

Venkatram Nalla

[1]*Centre for Disruptive Photonic Technologies, The Photonic Institute,
School of Physical and Mathematical Sciences, Nanyang Technological University, Singapore*

*Dedicated to Professor D N Rao for his significant contributions and pioneering works in the
fields of spectroscopy, optics, nonlinear optics and photonics*

Multiphoton absorption (MPA) is a nonlinear optical process that involves simultaneous absorption of two or more photons by a material to promote its ground state to an excited state. Multiphoton absorption promises many important applications such as multi-dimensional fluorescence imaging, three dimensional (3D) data storage. Physical and chemical properties of nanometer sized semiconductor materials change drastically due to quantum size effect. Combination of both multiphoton absorption and quantum effects will be an interesting study. Multiphoton absorption of semiconducting nanomaterials is an exciting phenomenon which promotes many important applications. This paper reviews multi-photon absorption properties of different kinds of semiconducting nanomaterials starting from chalcogenide-based nanomaterials to perovskites nanomaterials and their applications in various fields. © Anita Publications. All rights reserved.

**Keywords**: Multi-photon absorption, Nanomaterials, Semiconducting nanomaterials.

## 1 Introduction

In recent years, interest in the synthesis, characterization and application of semiconductor materials has grown markedly [1,2]. Due to control on their band gaps over wide spectral range, semiconductor nanoparticles have many potential applications in the fields of nanophotonics and optoelectronics. Semiconductor nanomaterials include chalcogenide-based nanomaterials (S, Se and Te derivatives) [3-5], silicon-based nanomaterials [6], metal oxides nanomaterials [7] and most recent perovskites nanomaterials [8,9]. Semiconductor nanomaterials are synthesized by many different methods which includes Colloidal synthesis [3,5], Laser-driven nanomaterials [10], Physical Vapour deposition / Chemical Vapour deposition [11] and Bio synthesis [12,13] etc. Semiconducting nanomaterials which are smaller than the Bohr exciton radius demonstrate unique optical properties due to the effects of three dimensional quantum confinement. Larger confinement potential for the smaller particle sizes (particle in a three dimensional spherical box model) leads to broadening in the band gap, which is inversely proportional to the particle size. Bohr exciton radius of famous semiconducting nanomaterials are well documented [14-16].

Multiphoton absorption is a nonlinear optical process that involves the simultaneous absorption of two or more photons to promote a material from its ground state to an excited state. This phenomenon has many important applications spanning the fields of Bio-imaging, multi-dimensional fluorescence imaging, photodynamic therapy, three dimensional (3D) data storage, optical limiting and 3D micromachining. Its advantages over single photon absorption include deeper sample penetration, better spatial selectivity and resolution, and reduced photochemical damage/decomposition and reduced light scattering at longer wavelengths [17,18]. The manifestation of quantum size effects in nonlinear optical processes depends on

*Corresponding author*
*e mail: nallavram@gmail.com* (Venkatram Nalla)
doi: https://doi.org/10.54955/AJP.30.6.2021.907-916

the electronic properties of the semiconductor, in particular, its band gap. A semiconductor nanoparticle is an example of a low-dimensional structure. The nanoparticle has rather a large number of atoms, but its size is comparable with characteristic dimensions describing the behavior of electrons and holes, thus creating an intermediate regime between molecules and bulk crystals [1]. Semiconducting materials have shown many interesting properties [3,19-22], because of the strong multiphoton absorption observed when their band gap is more than twice the photon energy (Eg > 2Ephoton), avoiding direct one-photon optical absorption. Three-photon absorption (3PA) and four-photon absorption (4PA) are particularly interesting as IR photons get converted to a blue or UV region. These multiphoton absorption processes play a major role in biological imaging, thereby increasing the resolution beyond the diffraction limit [23]. Multiphoton absorption in materials has also generated interest in laser direct writing [24]. Therefore, there is a need for materials that show high multiphoton absorption cross-sections. There are different methods being established to measure 3PA and 4PA spectra of wide-gap semiconductors, including well-known Z-scan measurements [25-29]. In this respect, femtosecond laser studies become important as the multiphoton absorption plays an important role at these time scales. Excitations at femtosecond timescales are important to overcome the contributions from free charge carrier absorption and thermal effects that accompany the nanosecond laser excitation. Multi-photon excited fluorescence is another important method to measure multiphoton absorption cross-sections [30]. This review is mainly focused on these two methods to measure multiphoton absorption cross-sections.

## 2 Result and Discussion

*Methods to measure multiphoton absorption cross-sections*

As multi photon absorption involves simultaneous absorption of multiple photons by a material to go to the higher excited states, optical properties of these materials such as transmission, reflection, absorption and emission will be nonlinear in nature. Currently, two main methods are followed to measure multiphoton absorption coefficients and multiphoton absorption cross-sections of different materials. First one is a direct method, which involves collecting input intensity dependent nonlinear transmission from the sample and fitting experimental data with nonlinear equations (by considering all the possible nonlinear process in the material) gives the absorption cross-section values. Second one is comparative/indirect method. This involves collecting multi-photon excited florescence intensity from the sample and well studied reference sample at the same experimental conditions. Multi photon absorption cross-section of the sample is estimated using ratio equation by comparing with reference values.

*a. Nonlinear Transmission (NLT) Method/ Z-scan method*

In a simple Z-scan measurements [31], laser beam is focused with a lens on the sample for which Rayleigh range should be larger than the sample thickness. The transmitted beam from the sample is focused onto the photodiode/detector with a lens. The transmitted data are recorded by scanning the sample across the focus using a computer controlled stepper motor and the data are collected at different Z- positions. Transmitted signals from photodiodes are connected to the lock-in amplifier / boxcar averager to increase signal to noise ratio, later analog-to-digital converters (ADC) are used to get final digital data. Simple Z-scan experimental setup is shown in Fig 1. Another well-established method is instead of moving sample in Z-direction, direct nonlinear transmission of the sample also can be measured by keeping sample at the focus (Z = 0) and collecting sample transmission by varying input powers with variable attenuator.

To estimate multi-photon absorption cross-sections of the sample, experimentally obtained data is fitted with theoretically simulated curves. Nonlinear equations are used to simulate theoretical curves, considering band diagrams of the sample, example of a band diagram is shown in Fig 2A. The nonlinear attenuation of the incident light is described by [5].

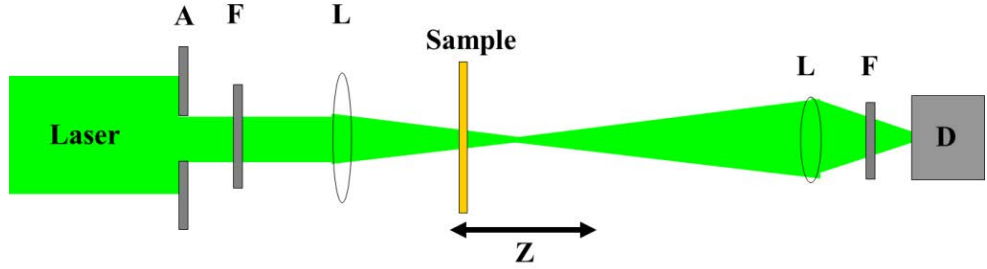

Fig 1. Schematic of the Z-scan set up for recording the multi-photon absorption. A - Aperture, S – Sample, F – Neutral Density Filters, D– Detectors and L – lens.

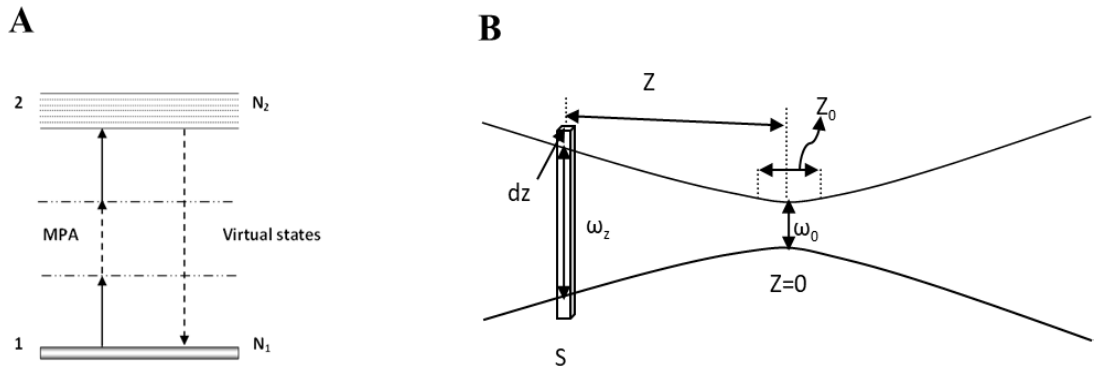

Fig 2. (A) Schematic of simple two level diagram, representing multi-photon absorption. (B) Details of focused beam for simulations.

$$\frac{dI}{dZ} = -\alpha_n I^n \tag{1}$$

where $\alpha_n$ is the *n*-photon absorption coefficient. As shown in Fig 2B focused beam diameter varies with Z position and intensity is power density which depends on beam diameter. Here, input intensity at the sample is varied by moving sample in Z-direction. By considering input intensity is Z dependent,

$$\frac{dI(Z)}{dZ} = -\alpha_n I^n(Z)$$

$$\int_{I_{in}}^{I_{out}} \frac{1}{I^n(Z)} dI(Z) = -\alpha_n \int_0^L dZ$$

$$-\alpha_n L = \left[\frac{I^{-n+1}}{-n+1}\right]_{I_{in}}^{I_{out}}$$

$$-\alpha_n L = \frac{1}{n-1}\left[\frac{1}{I_{in}^{n-1}} - \frac{1}{I_{out}^{n-1}}\right]$$

$$\alpha_n L = \frac{1}{(n-1)I_{in}^{n-1}}\left[1 - \frac{1}{T^{n-1}}\right]$$

where, $T = \frac{I_{out}}{I_{in}}$

$$T^{n-1} = \frac{1}{1 + \alpha_n L(n-1) I_{in}^{n-1}}$$

The transmitted light from sample can be written as:

$$T = \frac{1}{[1 + \alpha_n L(n-1) I_{in}^{n-1}]^{1/(n-1)}}$$

$$\frac{I_{in}}{\omega_0^2} = \frac{I_{00}}{\omega_Z^2}$$

$$I_{in} = \frac{I_{00}}{1 + \frac{Z^2}{Z_0^2}}$$

where, $\omega_Z^2 = \omega_0^2 \left(1 + \frac{Z^2}{Z_0^2}\right)$

$$T = \frac{1}{[1 + (n-1) \alpha_n L(I_{00}/(1 + (Z/Z_0)^2))^{n-1}]^{1/(n-1)}}$$

$$\Rightarrow T_{OA(nPA)} = \frac{1}{[1 + (n-1) \alpha_n L(I_{00}/(1 + (Z/Z_0)^2))^{n-1}]^{1/(n-1)}} \tag{2}$$

where $I_{00}$ is the peak intensity (at Z = 0), $I_{in}$ is laser intensity at position Z (Z - is the distance from focal point $Z_0$, $Z_0 = \pi\omega_0^2/\lambda$ is Rayleigh range, $\omega_0$ is the beam waist at the focal point (Z = 0), dZ is small slice of the sample, $T = I_{out}/I_{in}$ and $I_{out}$ is transmitted output intensity from the sample. L is sample effective path length. Effective path length for n-photon absorption is $L = [1 - \exp(-n\alpha_0 L_p)]/n\alpha_0$, where $L_p$ is sample physical path length and $\alpha_0$ is linear absorption coefficient at excitation wavelength, $\lambda$.

For isotropic samples, this *n*-photon absorption coefficient ($\alpha_n$) can be related to the *n*-photon absorption cross section ($\sigma_n$) by the formula:

$$\sigma_n = \frac{\alpha_n}{N_A d_0 10^{-3}} \left(\frac{hc}{\lambda}\right)^{n-1} \tag{3}$$

Here, $N_A$ is Avogadro's number and $d_0$ the molar concentration of the solution. $hc/\lambda$ is the photon energy with $h$, $c$ and $\lambda$ representing Planck's constant, the speed of light in vacuum, and the wavelength, respectively. The MPA cross-section is independent of sample concentration.

Typical Z-scan curve is shown in Fig 3A, nonlinear transmission of the sample is plotted for different Z positions. Sample exhibits maximum multi-photon absorption at the focus of the laser beam, where intensity is higher, and maximum dip in the transmission is observed at focal point Z = 0. From the best theoretical fit to the experimental Z-scan curves, multi-photon absorption coefficient ($\alpha_n$) and multi-photon absorption cross section ($\sigma_n$) were estimated [5,17]. Figure 3B is nonlinear transmission curve for CdS nanomaterials [3]. In nonlinear transmission measurements, sample position is fixed at focus (Z = 0) and transmission from the sample is measured for different input intensities.

*b. Multi-photon excited fluorescence method*

Multi-photon-absorption induced PL emissions of different materials are studied by exciting with femtosecond laser pulses. In this method, materials are excited by multi-photon absorption process (IR photons) and emissions are collected in visible region. Simple setup to study multi-photon-absorption induced PL emissions is shown in Fig 4, where incident laser pulses are focused by a lens onto the sample. The multi-photon-excited PL signal was collected in the perpendicular direction of the incident light using a collection system of two lenses; and then coupled into a spectrometer. In this method, multi-photon excited fluorescence spectra were obtained by comparison with a fluoresce in calibration of standard/ reference sample. Most commonly used reference samples are Rhodamine 6G, Rhodamine B and perylene crystals etc [30,32]. Multi-photon excited PL emission curve of ZnS nanomaterials is shown in Fig 5A, reference PL signal for Rhodamine 6G with the same setup is also shown. Order of nonlinearity (number of photons absorbed simultaneously) is determined from the slope of input intensity dependent PL emission intensity curve, as shown in Fig 5B. Here, slope of the curve determines the order of *n*-photon absorption.

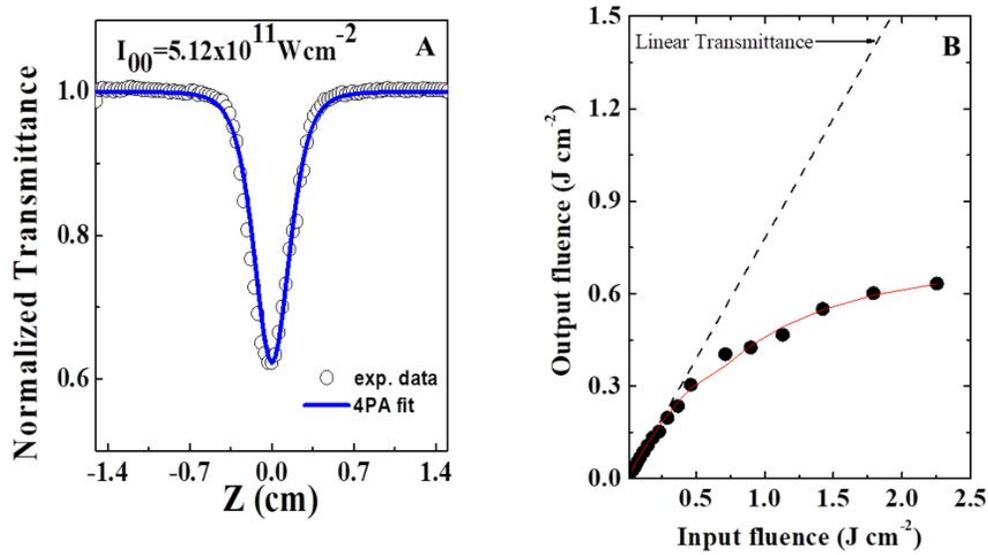

Fig 3. Nonlinear transmission curves due to multi-photon absorption. (A) Z-scan curve of CdSe nanocrystals exited at 800nm, 100fs laser pulses. (B) Nonlinear transmission of CdSe nanocrystals at different input fluences.

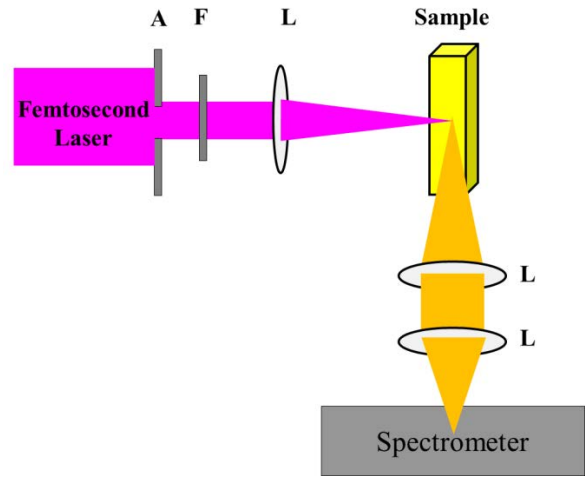

Fig 4. Schematic of the multi-photon excited florescence. The incident laser pulses are focused by a lens onto a sample. The PL signal is collected in the perpendicular direction of the incident light, by two collection lens system and then coupled into a spectrometer.

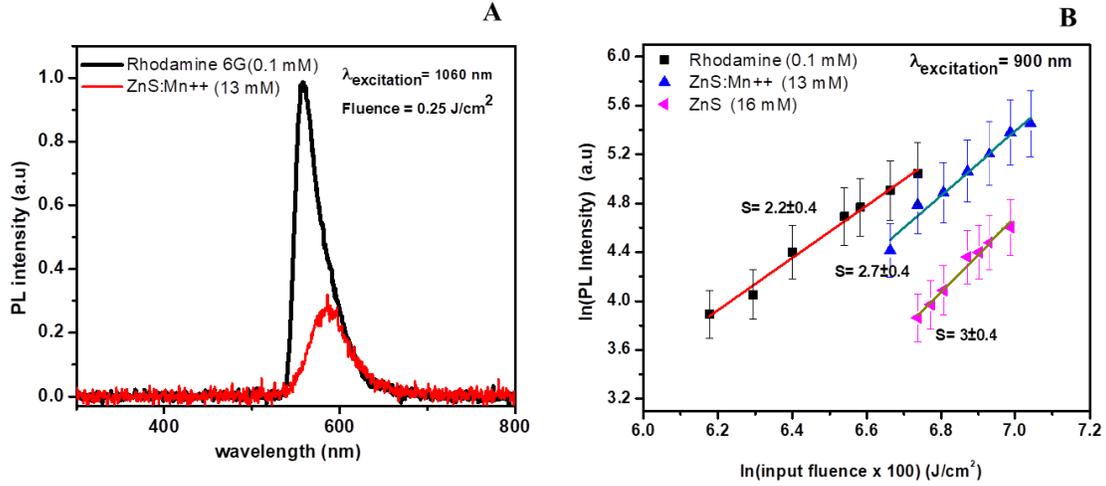

Fig 5. (A) Multi- photon excited PL emission signals for ZnS:Mn$^{++}$ quantum dots and Rhodamine 6G (excited at 1060 nm and Fluence 0.25 J/cm$^2$) signals are normalized to Rhodamine 6G PL intensity. B) Input laser power dependent multi-photon excited PL intensity in ZnS QD's and ZnS:Mn$^{++}$ QD's compared to Rhodamine 6G at 900nm excitation. Slope of these curves reveal the order of nonlinearity [30].

Quantification of multi-photon effective cross sections in multi-photon excited fluorescence method is done by comparing with absorption cross sections of well reported reference samples. Both, sample and reference sample are excited with the similar laser powers and at identical conditions. Fluorescence strength of the sample is estimated from the following equation. [33].

$$\Delta f_n = \eta \, \phi \, \sigma_n \, \rho \, ds \, dz \, I_r^n \tag{4}$$

where $\Delta f_n$ = n- photon-excited fluorescence strength, $\eta$ = fluorescence quantum yield, $\phi$ = fluorescence collection effciency of the experimental setup, $\sigma_n$ = n-photon absorption cross-section, $\rho$ = sample concentration, $ds \, dz$ = small volume of the focused laser beam, $I_r$ = laser intensity (nearly constant) at this small volume

$$F_2 = (\pi^{5/2}/4) \, \tau \, \phi \, \eta \, \sigma_n \rho \, I_{00}^n \, \omega_0^2 \, z_0 \tag{5}$$

where $\tau$ is the half width of the Gaussian laser pulse, $I_{00}$ is the peak intensity on the beam propagation axis, $\omega_0$ is the beam waist, and $z_0$ is the diffraction length.

The *n*-photon effective cross section value has been computed as,

$$\sigma_{n, \, sample} \, \eta_{sample} = \sigma_{n, \, reference} \, \eta_{reference} \, \frac{I_{sample} \, N_{reference}}{I_{reference} \, N_{sample}} \tag{6}$$

where the emission intensities were integrated over wavenumber, $I_{sample}$ is PL intensity of sample, $I_{reference}$ is PL intensity of reference and $N$ is molecules per unit volume (this is product of concentration and Avogadro's number $N_A$). As many reviews are available on multiphoton absorption cross-sections[17], therefore only recent works are being tabulated in Table 1.

## 3 Conclusion

Semiconducting nanomaterials represent one of the most exciting new developments for multi-photon applications. In the past few years, many attractive properties have been uncovered, such as extremely high emission quantum yield, broadband tunable wavelength emission, and giant multi-photon absorption/action cross-sections. New excitements are being added into the family of semiconducting nanomaterials through perovskite-based nanomaterials. These nanomaterials can be used in Bio-imaging, multi-dimensional fluorescence imaging, photodynamic therapy, three dimensional (3D) data storage, optical limiting and 3D micromachining. There is a lot of prospects need to be explored, especially applications of semiconducting nanomaterials in different fields. Table 1 summarizes the some of the finest semiconducting nanomaterials with high multi-photon absorption cross sections.

Table 1. Some of the semiconducting nanomaterials with high multi-photon absorption cross-sections (QDs- quantum dots, NCs- nanocrystals, NRs-nanorods). For 2- photon absorption cross-sections GM = $10^{-50}$ cm$^4$sphoton$^{-1}$.

| Material | Order of photon absorption | Multi-photon absorption cross-section | Excitation Wavelength (nm) | Reference |
|---|---|---|---|---|
| CsPbI$_3$ QDs (Halide perovskite) | 2-photon | $2.1 \times 10^6$ GM | 800 | 34 |
| | 3-photon | $1.1 \times 10^{-73}$ cm$^6$s$^2$photon$^{-2}$ | 1400-1500 | |
| CsPbCl$_3$ (Halide perovskite) | 2-photon | $3.8 \times 10^4$ GM | 800 | |
| Mn$^{2+}$-doped ZnS QDs | 2-photon | $2.6 \times 10^2$ GM | 1200 | 30 |
| CdSe QDs | 3-photon | $2.8 \times 10^{-77}$ cm$^6$s$^2$photon$^{-2}$ | 1300 | 35, 36 |
| ZnSe/ZnS-II QDs | 3-photon | $2.4 \times 10^{-75}$ cm$^6$s$^2$photon$^{-2}$ | | 37 |
| CdSe/CdS (core–shell QDs) | 3-photon | $4.3 \times 10^{-78}$ cm$^6$s$^2$photon$^{-2}$ | 1300 | 4 |
| CdSe/CdS/ZnS (core–shell QDs) | 3-photon | $2.8 \times 10^{-77}$ cm$^6$s$^2$photon$^{-2}$ | | |
| MAPbBr$_3$/(OA)$_2$PbBr$_4$ (core–shell perovskite NCs) | 2-photon | $3 \times 10^6$ GM | 675-1000 | 38 |
| | 3-photon | $2.5 \times 10^{-74}$ cm$^6$s$^2$photon$^{-2}$ | 1050-1500 | |
| | 4-photon | $2.1 \times 10^{-105}$ cm$^8$s$^3$photon$^{-3}$ | 1550-2000 | |
| | 5-photon | $2.9 \times 10^{-137}$ cm$^{10}$s$^4$photon$^{-4}$ | 2050-2300 | |
| Cs$_4$PbBr$_6$ perovskite 0D-films | 2-photon | $\sim 10^6$ GM | 500-800 | 39 |
| | 3-photon | $\sim 10^{-73}$ cm$^6$s$^2$photon$^{-2}$ | 900-1200 | |
| | 4-photon | $\sim 10^{-100}$ cm$^8$s$^3$photon$^{-3}$ | 1300-1500 | |
| CsPbBr$_3$ NRs | 3-photon | $3.6 \times 10^{-75}$ cm$^6$s$^2$photon$^{-2}$ | 1300 | 40 |
| ZJU-28⊃MAPbBr$_3$ (MOF-Perovskite QDs) | 3-photon | $2.7 \times 10^{-74}$ cm$^6$s$^2$photon$^{-2}$ | 1440 | 41 |
| | 4-photon | $5.3 \times 10^{-105}$ cm$^8$s$^3$photon$^{-3}$ | 1880 | |
| | 5-photon | $10.4 \times 10^{-136}$ cm$^{10}$s$^4$photon$^{-4}$ | 2100 | |
| CsPbBr$_{2.7}$I$_{0.3}$ (Perovskite 2D nanoplates) | 3-photon | $2.3 \times 10^{-74}$ cm$^6$s$^2$photon$^{-2}$ | 1300 | 42 |
| | 4-photon | $2.06 \times 10^{-104}$ cm$^8$s$^3$photon$^{-3}$ | 1600 | |
| | 5-photon | $1.5 \times 10^{-136}$ cm$^{10}$s$^4$photon$^{-4}$ | 2200 | |
| CsPbCl$_3$ (Microcrystals) | 3-photon | $\beta_3 = 0.089$ cm$^3$/GW$^2$ | 1200 | 43 |
| | 4-photon | $\beta_4 = 1.1 \times 10^{-4}$ cm$^5$/GW$^3$ | 1600 | |
| | 5-photon | $\beta_5 = 2.3 \times 10^{-7}$ cm$^7$/GW$^4$ | 1800 | |
| | 6-photon | $\beta_6 = 1.1 \times 10^{-10}$ cm$^9$/GW$^5$ | 2200 | |
| Mn-doped CsPbCl$_3$ (Nanocrystals) | 3-photon | $1.49 \times 10^{-76}$ cm$^6$s$^2$photon$^{-2}$ | 1300 | 44 |
| Mn-doped CsPbCl$_3$ (2D nanoplates) | 3-photon | $6.54 \times 10^{-76}$ cm$^6$s$^2$photon$^{-2}$ | 1300 | |

**References**

1. Alivisatos A P, Semiconductor Clusters, Nanocrystals, and Quantum Dots, *Science*, 271(1996)933–937.
2. Alivisatos A P, Perspectives on the physical chemistry of semiconductor nanocrystals, *J Phys Chem*, 100(1996) 13226–13239.
3. Venkatram N, Rao D N, Akundi M A, Nonlinear absorption, scattering and optical limiting studies of CdS nanoparticles, *Opt Express*, 13(2005)867–872.
4. Yue W, Swee L K, Duong T V, Rui C, Venkatram N, Yuan G, Tingchao H, Volkan D H, Handong S, Blue Liquid Lasers from Solution of CdZnS/ZnS Ternary Alloy Quantum Dots with Quasi-Continuous Pumping, *Adv Mater*, 27(2015)169–175.
5. Venkatram N, Sathyavathi R, Rao D N, Size dependent multiphoton absorption and refraction of CdSe nanoparticles, *Opt Express*, 15(2007)12258–12263.
6. Barua S, Gogoi S, Khan R, Karak N, Chapter 8 - Silicon-Based Nanomaterials and Their Polymer Nanocomposites, In, Nanomaterials and Polymer Nanocomposites (2019), 261-305; doi. 10.1016/B978-0-12-814615-6.00008-4.
7. Pan S, Jeevanandam J, Acquah C, Tan K X, Udenigwe C C, Danquah M K, Chapter 25 - Drug delivery systems for cardiovascular ailments, (eds): Eric Chappel, In Developments in Biomedical Engineering and Bioelectronics, (Academic Press), 2021, pp 567-599.
8. Yue W, Xiaoming L, Venkatram N, Haibo Z, Handong S, Solution-Processed Low Threshold Vertical Cavity Surface Emitting Lasers from All-Inorganic Perovskite Nanocrystals, *Adv Funct Mater*, 27(2017)1605088; doi/abs/10.1002/adfm.201605088.
9. Yue W, Dejian Y, Zeng W, Xiaoming L, Xiaoxuan C, Venkatram N, Haibo Z, Handong S, Solution-Grown CsPbBr3/Cs4PbBr6 Perovskite Nanocomposites: Toward Temperature-Insensitive Optical Gain, *Small*, 13(2017)1701587; doi/abs/10.1002/smll.201701587
10. Shujun Wang, Lihong Gao, Chapter 7 - Laser-driven nanomaterials and laser-enabled nanofabrication for industrial applications, Eds: Sabu Thomas, Yves Grohens, Yasir Beeran Pottathara, In Micro and Nano Technologies, Industrial Applications of Nanomaterials, (Elsevier), 2019, pp 181-203. https://www.sciencedirect.com/science/article/pii/B9780128157497000074
11. Reilly C E, Keller S, Nakamura S, DenBaars S P, Metalorganic chemical vapor deposition of InN quantum dots and nanostructures, *Light Sci Appl*, 10(2021)150; doi.org/10.1038/s41377-021-00593-8.
12. Dameron C T, Reese R N, Mehra R K, Kortan A R, Carroll P J, Steigerwald M L, Brus L E, Winge D R, Biosynthesis of cadmium sulphide quantum semiconductor crystallites, *Nature*, 338(1989)596–597.
13. Salouti M, Faghri Zonooz N. Biosynthesis of Metal and Semiconductor Nanoparticles, Scale-Up, and Their Applications. In: Ghorbanpour M, Manika K, Varma A. (eds) Nanoscience and Plant–Soil Systems. Soil Biology, vol 48. (Springer), 2017, doi.org/10.1007/978-3-319-46835-8_2.
14. Guzelturk B, Martinez P L H, Zhang Q, Xiong Q, Sun H, Sun X W, Govorov A O, Demir H V, Excitonics of semiconductor quantum dots and wires for lighting and displays, *Laser Photonics Rev*, 8(2014)73–93.
15. Wang Q, Liu X-D, Qiu Y-H, Chen K, Zhou L, Wang Q-Q, Quantum confinement effect and exciton binding energy of layered perovskite nanoplatelets, *AIP Adv*, 8(2018)025108; doi:10.1063/1.5020836.
16. Li X M, Cao F, Yu D J, Chen J, Sun Z G, Shen Y L, Zhu Y, Wang L, Wei Y, Wu Y, Zeng H B, All Inorganic Halide Perovskites Nanosystem: Synthesis, Structural Features, Optical Properties and Optoelectronic Applications, *Small*, 13(2017)1603996; doi/10.1002/smll.201603996.
17. He G S, Tan L.-S, Zheng Q, Prasad P N, Multiphoton Absorbing Materials: Molecular Designs, Characterizations, and Applications, *Chem Rev*, 108(2008)1245–1330.
18. Guo L, Wong M S, Multiphoton Excited Fluorescent Materials for Frequency Upconversion Emission and Fluorescent Probes, *Adv Mater*, 26 (2014)5400–5428.

19. Villeneuve A, Yang C C, Wigley P G J, Stegeman G I, Aitchison J S, Ironside C N, Ultrafast all-optical switching in semiconductor nonlinear directional couplers at half the band gap, *Appl Phys Lett*, 61(1992)147–149.
20. Yang C C, Villeneuve A, Stegeman G I, Aitchison J S, Effects of three-photon absorption on nonlinear directional coupling, *Opt Lett*, 17(1992)710-712.
21. Venkatram N, Kumar R S S, Rao D N, Nonlinear absorption and scattering properties of cadmium sulphide nanocrystals with its application as a potential optical limiter, *J Appl Phys*, 100(2006)074309; doi/10.1063/1.2354417.
22. Morgan R A, Park S H, Koch S W, Peyghambarian N, Experimental studies of the non-linear optical properties of cadmium selenide quantum-confined microcrystallites, *Semicond Sci Technol*, 5(1990)544; doi.org/10.1088/0268-1242/5/6/014.
23. Chon J W M, Gu M, Bullen C, Mulvaney P, Three-photon excited band edge and trap emission of CdS semiconductor nanocrystals, *Appl Phys Lett*, 84 (2004)4472–4474.
24. Deubel M, von Freymann G, Wegener M, Pereira S, Busch K, Soukoulis C M, Direct laser writing of three-dimensional photonic-crystal templates for telecommunications, *Nature Mater*, 3(2004)444–447.
25. He J, Qu Y, Li H, Mi J, Ji W, Three-photon absorption in ZnO and ZnS crystals, *Opt Express*, 13(2005)9235–9247.
26. Sutherland R L with contributions by McLean D G and Kirkpatrick S, Handbook of Nonlinear Optics, 2nd Edn, Revised and Expanded, (New York, NY: Marcel Dekker), 2003:. https://www.taylorfrancis.com/books/edit/10.1201/9780203912539/handbook-nonlinear-optics-richard-sutherland.
27. Bindra K S, Kar A K, "Role of femtosecond pulses in distinguishing third- and fifth-order nonlinearity for semiconductor-doped glasses, *Appl Phys Lett*, 79( 2001)3761–3763.
28. He G S, Zheng Q, Baev A, Prasad P N, Saturation of multiphoton absorption upon strong and ultrafast infrared laser excitation, *J Appl Phys*, 101(2007)083108; doi/10.1063/1.2719286.
29. Pacebutas V, Krotkus A, Suski T, Perlin P, Leszczynski M, Photoconductive Z-scan measurement of multiphoton absorption in GaN, *J Appl Phys*, 92(2002)6930–6932.
30. Subha R, Nalla V, Yu J H, Jun S W, Shin K, Hyeon T, Vijayan C, Ji W, Efficient hotoluminescence of $Mn^{2+}$-doped ZnS Quantum Dots Excited by Two-photon Absorption in Near-IR Window II, *Phys Chem C*, 117(2013)20905–20911.
31. Sheik-Bahae M, Said A A, Wei T H, Hagan D J, Van Stryland E W, Sensitive measurement of optical nonlinearities using a single beam, *IEEE J Quantum Electron*, 26 (1990)760–769.
32. Makarov N S, Drobizhev M, Rebane A, Two-photon absorption standards in the 550-1600 nm excitation wavelength range, *Opt Express*, 16(2008)4029–4047.
33. Medishetty R, Nalla V, Nemec L, Henke S, Mayer D, Sun H, Reuter K, Fischer R A, A New Class of Lasing Materials: Intrinsic Stimulated Emission from Nonlinear Optically Active Metal–Organic Frameworks, *Adv Mater*, 29(2017)1605637; doi/abs/10.1002/adma.201605637.
34. Pramanik A, Gates K, Gao Y, Begum S, Ray P C, Several Orders-of-Magnitude Enhancement of Multiphoton Absorption Property for CsPbX3 Perovskite Quantum Dots by Manipulating Halide Stoichiometry, *J Phys Chem C*, 123(2019)5150–5156.
35. Feng X, Ang Y L, He J, Beh C W J, Xu H, Chin W S, Ji W, Three-photon absorption in semiconductor quantum dots: experiment, *Opt express*,16(2008)6999–7005.
36. He G S, Yong K-T, Zheng Q, Sahoo Y, Baev A, Ryasnyanskiy A I, Prasad P N, Multi-photon excitation properties of CdSe quantum dots solutions and optical limiting behavior in infrared range, *Opt Express*, 15(2007)12818–12833.
37. Lad A D, Kiran P P, Kumar G R, Mahamuni S, Three-photon absorption in ZnSe and ZnSe/ZnS quantum dots, *Appl Phys Lett*, 90 (2007)133113;doi.org/10.1063/1.2714994.
38. Chen W, Bhaumik S, Veldhuis S A, Xing G, Xu Q, Grätzel M, Mhaisalkar S, Mathews N, Sum T C, Giant five-photon absorption from multidimensional core-shell halide perovskite colloidal nanocrystals, *Nat Commun*, 8(2017)15198; doi.org/10.1038/ncomms15198.
39. Krishnakanth K N, Seth S, Samanta A, Venugopal Rao S V, Broadband ultrafast nonlinear optical studies revealing exciting multi-photon absorption coefficients in phase pure zero-dimensional $Cs_4PbBr_6$ perovskite films, *Nanoscale*, 11(2019)945-954.

40. Li J, Jing Q, Xiao S, Gao Y, Wang Y, Zhang W, Sun X W, Wang K, He T, Spectral Dynamics and Multiphoton Absorption Properties of All-Inorganic Perovskite Nanorods, *J Phys Chem Lett*, 11(2020)4817–4825.
41. He H, Cui Y, Li B, Wang B, Jin C, Yu J, Yao L, Yang Y, Chen B, Qian G, Confinement of perovskite-QDs within a single MOF crystal for significantly enhanced multiphoton excited luminescence, *Adv Mater*, 31(2019)1806897; doi/10.1002/adma.201806897.
42. Li J, Zhao F, Xiao S, Cheng J, Qiu X, Lin X, Chen R, He T, Giant two-to five-photon absorption in $CsPbBr_{2.7}I_{0.3}$ two-dimensional nanoplatelets, *Opt Lett*, 44(2019)3873–3876.
43. Yang D, Chu S, Wang Y, Siu C K, Pan S, Yu S F, Frequency upconverted amplified spontaneous emission and lasing from inorganic perovskite under simultaneous six-photon absorption, *Opt Lett*, 43(2018)2066–2069.
44. He T, Li J, Qiu X, Xiao S, Lin X, Superior multiphoton absorption properties in colloidal Mn-doped $CsPbCl_3$ two-dimensional nanoplatelets, *Photonics Res*, 6(2018)1021–1027.

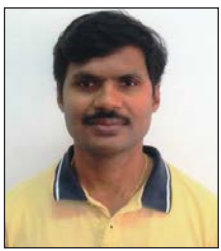

Venkatram Nalla is one of the Prof D Narayana Rao's Ph D students who graduated in 2008 (from University of Hyderabad). Later, he worked as Post Doc researcher at National University of Singapore and Nanyang Technological University, Singapore. He was also visiting researcher at ORC, University of Southampton, UK.

Venkatram Nalla authored more than 50 international journal publications such as *Nature Communications*, *Advanced Materials*, *Angewandte Chemie*, *Nano Letters*, *Advanced Functional Materials* and *Optics express* etc. His recent research area includes fields of Nano-photonics, Quantum dots, Micro lasers, Metamaterials, sub-10fs ultrafast optical switching and 100THz optical switches. He is currently working on Semiconductor Failure Analysis tools. He is reviewer of Asian J Physics (AJP).